# A review on non-Hermitian skin effect

Xiujuan Zhang[1,†], Tian Zhang[1], Ming-Hui Lu[1,2,3,†] and Yan-Feng Chen[1,3]

[1]*National Laboratory of Solid State Microstructures and Department of Materials Science and Engineering, Nanjing University, Nanjing 210093, China*

[2]*Jiangsu Key Laboratory of Artificial Functional Materials, Nanjing 210093, China*

[3]*Collaborative Innovation Center of Advanced Microstructures, Nanjing University, Nanjing 210093, China*

[†]Correspondence should be addressed to xiujuanzhang@nju.edu.cn (XJZ) or luminghui@nju.edu.cn (MHL).

## Abstract

The past decades have witnessed the flourishing of non-Hermitian physics in non-conservative systems, leading to unprecedented phenomena of unidirectional invisibility, enhanced sensitivity and more recently the novel topological features such as bulk Fermi arcs. Among them, growing efforts have been invested to an intriguing phenomenon, known as the non-Hermitian skin effect (NHSE). Here, we review the recent progress in this emerging field. By starting from the one-dimensional (1D) case, the fundamental concepts of NHSE, its minimal model, the physical meanings and consequences are elaborated in details. In particular, we discuss the NHSE enriched by lattice symmetries, which gives rise to unique non-Hermitian topological properties with revised bulk-boundary correspondence (BBC) and new definitions of topological invariants. Then we extend the discussions to two and higher dimensions, where dimensional surprises enable even more versatile NHSE phenomena. Extensions of NHSE assisted with extra degrees of freedom such as long-range coupling, pseudospins, magnetism, non-linearity and crystal defects are also reviewed. This **is followed by the contemporary experimental progress for NHSE. Finally, we provide the outlooks to possible future directions and developments.**

## 1. Introduction

The Hermiticity of a Hamiltonian guarantees its real eigenvalues and orthogonal eigenstates, which reflects the dynamics and shapes the physical reality of an isolated quantum system with conversed energy and probability. When encountering with non-conservative systems, however, gain/loss of the energy and probability leads to the breakdown of Hermiticity and requires non-Hermitian descriptions [1-3]. It has been shown the non-Hermitian considerations not only provide suitable descriptions to the open systems, but also bring novel physics and unprecedented phenomena and applications, as found in a variety of physical realms including open quantum systems [4], electronic systems with interactions [5], and classical systems with gain or loss [6-20]. For instance, mathematics suggests that the real eigenvalues do not necessarily rely on the Hermiticity, but can also be realized if the parity-time (PT) symmetry is satisfied as a form of pseudo-Hermiticity [1,3]. Such a mathematical prediction later on became a physical reality in non-Hermitian systems where balancing material gain and loss can lead to PT symmetry [9], associated with which, the eigenvalues of the non-Hermitian Hamiltonian are pure real. Once the balance between gain and loss is broken, the eigenvalues become complex and a transition point emerges in the parameter space, known as the exceptional point (EP). The EP signals the transition from the PT-symmetric phase to PT-broken phase. At the EP, both the eigenvalues and the eigenvectors coalesce, enabling exotic features such as unidirectional invisibility [11] and enhanced sensitivity [14].

On another front, due to the ubiquitous existence of non-Hermiticity in both naturally-occurring materials and artificial materials, its interaction with periodic/aperiodic lattices has inspired a surge of research. It was shown that by engineering a photonic crystal plate that radiates energy to its surrounding, a pair of EPs can be created in the band structures in the momentum space, connected to each other by an open-ended bulk Fermi arc [15], in direct contrast to the isofrequency contours with closed loops in Hermitian systems. This phenomenon was found to be closely related to the band topology interplayed with non-Hermiticity. Along this direction, extensive research effort has been made to explore novel non-Hermitian band physics and associated intriguing phenomena [21-25]. Therein, a particularly interesting branch has drawn great attention recently, known as the NHSE [26], which describes the phenomenon that driven by non-Hermiticity, eigenstates of a lattice with open boundaries exhibit localized behaviors, drastically different from the extended Bloch waves in Hermitian systems. This phenomenon can be dated back to late 1990s where Hatano and Nelson [27-28] proposed a 1D disordered tight-binding lattice with nearest-neighbor nonreciprocal hopping and showed that this nonreciprocity-induced non-Hermiticity can prevent Anderson localization, opening up a mobility region featuring unidirectional transport.

Very recently, similar phenomenon was also observed during the investigation of the topological properties of non-Hermitian systems. In Ref. [29], it was found driving by nonreciprocal gauging, the topological edge states become defective and only one of them is stable while the other is not and dynamically transits into the stable one. Eventually, two edge states become identical. Similar behaviors were later discussed in another literature where not only defective edge states were reported, but also all the bulk states were found to change from extended states to exponentially localized states only by a change of the boundary condition from periodic to open [30]. These phenomena are counterintuitive to the common rules in Hermitian topological systems and challenge the well-developed BBC, therefore igniting tremendous research interest. Some seminal works include revisiting the Hatano-Nelson (HN) model to realize versatile directional transports [31-35], interpreting NHSE from different points of view [36-48], developing new theories, methods and material designs to describe the NHSE-related topological properties and the associated non-Hermitian BBC [26,49-89], extending NHSE from simple 1D studies to two and higher dimensions [78,90-102], exploring novel NHSE features enriched by various degrees of freedom [103-120], and experimentally demonstrating the NHSE [114,117,121-130].

Following these progresses, in this review, we start with the introduction of NHSE in the minimal 1D HN model (without any symmetry), its theoretical and physical interpretations, then move to the non-Hermitian systems with symmetries where the interplay between band topology (enabled by various symmetries) and NHSE leads to re-defined BBC and novel non-Hermitian topological properties. We next turn our discussions to higher dimensions with a focus on the 2D systems where dimension surprisingly gives rise to even more interesting phenomena such as higher-order NHSE and non-Bloch PT symmetry breaking without threshold. This is followed by an exhibition of the NHSE enriched by various degrees of freedom, such as long-range coupling, pseudospins, magnetism, non-linearity and crystal defects. The limited experimental progresses will also be included. Finally, we provide the conclusions and our perspectives on future possible directions.

Here, we mention that because the existing theories and methods for the studies of NSHE are mostly well-defined and well-established in one dimension while the other topics including NHSE in higher dimensions, carrying extra physical degrees of freedom, and the experimental demonstrations are still prosperously ongoing, our main focus will be on the 1D NHSE phenomena.

## 2. 1D NHSE

### 2.1 The HN model as a minimal model for NHSE

As discussed above, the HN model was originally proposed to study the non-Hermitian delocalization and later on was revisited for its unique unidirectional transports enabled by the newly identified NHSE. In this section, we introduce the fundamentals of the NHSE based on a clean HN model without disorder, where the Hamiltonian descriptions, the manifestation of the NHSE and its consequences on eigenstates are provided. The HN model is a single particle, spinless, prototypical 1D non-Hermitian lattice with nonreciprocal hopping, as schematically illustrated in Figure 1(a), where the left and right hopping is denoted respectively by $t + \gamma$ and $t - \gamma$ with $\gamma$ characterizing the strength of nonreciprocity and therefore the strength of non-Hermiticity. Without considering the disorders and onsite energy potential, the Hamiltonian under periodic boundary condition (PBC) reads

$$H_{\text{HN}}(k) = (t+\gamma)e^{ik} + (t-\gamma)e^{-ik}, \tag{1}$$

where $k$ represents the Bloch wave vector and the lattice constant is taken as 1 throughout this review unless specified otherwise. Figure 1(b) shows the single band structure when $k$ runs over the 1D Brillouin zone (BZ). As expected, in such a non-Hermitian system, the eigenenergy $E$ becomes complex, with the real part still symmetric with respect to $k$ while the imaginary part exhibiting an asymmetric property. In fact, this asymmetric property is nothing but an exact indicator of the nonreciprocal nature of the system, suggesting an asymmetric transport behavior. That is, the probability of the left hopping does not equal to that of the right hopping. In an extreme case where $t = \gamma$, the system only allows for the left hopping while the right hopping is forbidden. Consequently, upon open boundary conditions (OBCs), particles accumulate toward the left boundary, as shown by the eigenmode distributions in Figure 1(c-d) for two exemplified cases with $t = \gamma = 1$ and $t = 1, \gamma = 0.4$. Because the accumulation and localization behaviors share a phenomenological similarity to the skin accumulation of an alternating AC electric current in a conductor, this non-Hermitian phenomenon is dubbed NHSE and the localized modes are referred to as skin modes [26].

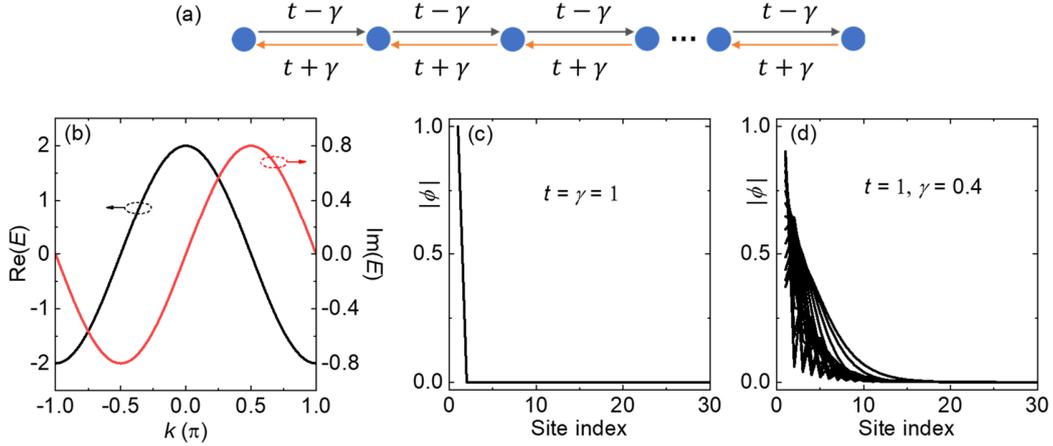

Figure 1. (a) The HN model. (b) The energy band under PBC. (c-d) Mode profiles of all the eigenstates under OBCs. The total number of sites (i.e., the chain length) is taken as $N = 30$.

The NHSE profoundly changes the Bloch band theory. As discussed above, associated with NHSE, the eigenstates of the systems loss their extendedness in the Bloch scope and become exponentially localized to lattice boundary. Such an intriguing property calls for revisiting the band theory in non-Hermitian systems and therefore has inspired novel research especially on nonreciprocal band physics, such as non-Hermitian topological properties which will be discussed later. In addition to the significances on band physics, the nonreciprocity-induced non-Hermiticity also holds great promises for directional transports and beyond. To list a few, in Refs. [32,34], Longhi *et al.* proposed and analyzed the unique unidirectional and bidirectional transports by nonreciprocal gauging in HN chains. Using a similar technique, the same author also suggested a method for excitation transfer [35], providing a shortcut to the adiabatic process while the latter usually requires long device length and/or long evolution time. In another proposal, it has been shown that under the threading of an imaginary vector potential (which can be effectively realized by nonreciprocal gauging), the optical gradient force can be enhanced by an order of magnitude, offering novel methods for actuation of microscopic objects using optical forces [131].

**2.2 Interpretations of NHSE**

After introducing the NHSE and its basics phenomena, next, we review the theories and methods to interpret NHSE from different points of view. There are three typical theories. The first one considers the NHSE as the effect of imaginary gauge fields, commonly used in early interpretations of non-Hermitian delocalization and directional transports in HN models [27, 28, 31-35]. It is known that threaded by a real gauge field, the particles accumulate a nontrivial real phase factor along their trajectory, which contributes to interferences that eventually lead to unidirectional transport, as in quantum Hall effect [132] and its various classical analogues [133, 134]. In the case of imaginary gauge fields, however, the particles accumulate an imaginary phase factor along their trajectory, which, instead of contributing to interferences, reduces or increases the particles' amplitudes along the direction in which they hop [31,131]. Essentially, this process can be physically understood as directional attenuation or amplification, which, upon OBCs, leads to exponential localizations as shown in Figure 1(c-d).

The second theory is related with the EPs. As discussed in the introduction section, the EPs refer to the singularities in non-Hermitian systems, appearing either in the parameter space [9] or the momentum space [15]. In these systems, the non-Hermiticity usually comes from the material gain and loss. It has been shown that at the EPs, both the eigenvalues and eigenvectors coalesce [1]. A regular EP represents the coalescence of two states [as schematically illustrated in Figure 2(a)], which generally induces a square root singularity, making it more sensitive to the parameter change compared to a Hermitian degenerate point and can be potentially used as a novel mechanism for sensors [14,135]. Higher-order EPs are coalescences of multiple states [see Figure 2(b)], carrying higher-order singularities, which make them even more sensitive than the first-order EPs [136].

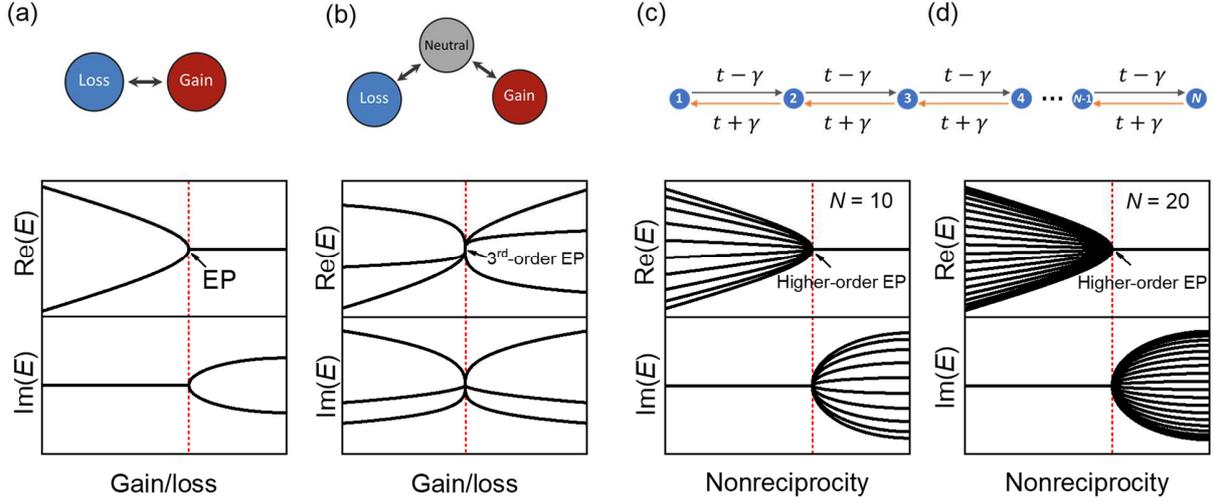

Figure 2. Sketches of (a) a regular and (b) a third-order EPs in gain/loss-induced two-level and three-level non-Hermitian systems. (c-d) Higher-order EPs emerging in the HN model with OBCs.

Contrary to the high sensitivity of EPs in the gain/loss-induced non-Hermitian systems, recent studies have shown that EPs in nonreciprocity-induced non-Hermitian systems have strong fingerprints that can effectively pervade over a wide parameter regime and are possibly associated with the emergence of NHSE [30,36]. Specifically, it was found that in nonreciprocity-induced non-Hermitian open systems, high-order EPs appear with algebraic multiplicities scaling with the system size, whereas the geometric multiplicities remain as 1, as depicted in Figure 2(c-d). This indicates that at the EPs, all the bulk states coalesce at one state, which was found exponentially localized at lattice boundary under OBCs, corresponding to the non-Hermitian skin mode. Interestingly, these multiple coalescences not only affect the eigenstates near the EPs, but also have a global effect on the entire eigen-spectrum, leading to an extensive amount of bulk states transiting into skin modes [see Figure 1(c-d)]. Only far from the EPs, the NHSE becomes less obvious, i.e., the localization length becomes longer and the eigenstates restore certain extendedness. The special EPs induced by nonreciprocity have recently been further investigated and found to exhibit novel behaviors such as the unique existence under OBCs [137] and vanishing threshold for PT phase transition in two dimensions [138]. These features typically cannot be observed in traditional non-Hermitian systems with gain and loss.

The third theory characterizing NHSE is based on the evaluation of the winding numbers of the eigenenergy. Studying topological phases of matter has been a hot research topic in condensed matter physics, whose interesting phenomena are generally enabled by the nontrivial topological properties of the wave functions, *not* the eigenenergy, of the occupied bands (or in classical wave systems the group of bands below a band gap) [139-142]. When extending the concept of topological phases of matter to non-Hermitian systems, one can immediately identify a novel type of topological number given solely by eigenenergy. That is, by phase-winding the eigenenergy, *not* the wave functions, of a non-Hermitian Hamiltonian over the BZ, a nontrivial winding number emerges, which constitutes a topological object unique to the non-Hermitian systems [143]. For instance, consider the HN model shown in Figure 1(a) and replot its energy band in Figure 1(b) in the complex plane, as presented in Figure 3(a). It is shown that the eigenenergy forms a closed loop, with the direction of $k$-winding from $-\pi$ to $\pi$ indicated by the arrow. For any base point $E_b$ inside the loop, one has a nonzero winding number following the definition [44]

$$w := \frac{1}{2\pi} \int_{-\pi}^{\pi} \frac{d}{dk} \arg[H(k) - E_b] \, dk. \qquad (2)$$

Essentially, $w$ gives the number of times the complex eigenenergy encircles the base point $E_b$. For the HN model shown in Figure 1(a), the eigenenergy encircles one time the base points inside its loop, as depicted in Figure 3(a). Taking the convention that clockwise (anticlockwise) circling gives negative (positive) winding number, one has $w = 1$ for the present example. As a comparison, the eigenenergy of the Hermitian counterpart is also plotted, which constitutes a straight line [see Figure 3(b)], corresponding to a vanishing winding number.

The topological interpretation of such a unique property of the complex eigenenergy in the non-Hermitian systems is elucidated in the following. According to the conventional BBC, for each base point $E_b$, the nonzero winding number $w$ indicates $|w|$ independent edge states with energy $E_b$. Since $E_b$ can be any point inside the loop of the eigenenergy of $H(k)$ under PBC, there are an infinite number of $E_b$, indicating infinite edge states. These edge states were shown to be manifested in the semi-infinite truncations of the non-Hermitian systems with only one boundary [43,144]. Upon full OBCs, i.e., the left and right open boundaries appear in a pair, the condition for infinite edge states breaks down. While some of the eigenstates are forbidden due to the finite length, the spectrum under the semi-infinite condition still includes the spectrum under the full OBCs as the former is the extrapolation of the latter under $L \to \infty$, where $L$ denotes the chain length [43]. As a result, the eigenstates under the full OBCs still preserve the behaviors of the edge states in the semi-infinite systems and are exponentially localized, corresponding to the non-Hermitian skin modes. It should be pointed out that even though their emergence has topological origin as discussed, the skin modes are not topological [34,36,92]. In fact, the eigenenergy under full OBCs no longer forms any loops nor encircles any base points, but instead collapses to arcs [see Figure 3(c) for an example of the HN model under full OBCs]. Accordingly, the winding number under full OBCs is either zero or ill-defined [144]. Nevertheless, the well-defined, nonzero $w$ under PBCs is still a valid indicator, associated with which, the NHSE always manifests under OBCs [43, 44].

Recently, the correspondence between the nonzero winding of the eigenenergy under PBCs and the emergence of non-Hermitian skin modes under OBCs was further explored and their relations were rigorously established [44], via a novel concept of the generalized Brillouin zone (GBZ) [26, 57, 70]. It has been noticed in multiple occasions that due to the presence of NHSE, the eigenspectrum of certain non-Hermitian systems under OBCs drastically deviates from that under PBCs [26, 49, 90]. To tackle this issue, the concept of GBZ was proposed in the scope of non-Bloch band theory [26]. This theory takes a complex wave vector $k' = k - i \ln r$ (it is in this form for the sake of concise formulism, with $r$ some real values) to replace the initial real wave vector $k$ and defines the GBZ based on $k'$ so that the non-Hermitian dynamics can be characterized as the non-Bloch wave propagation governed by the new phase factor $\beta = e^{ik'} = re^{ik}$.

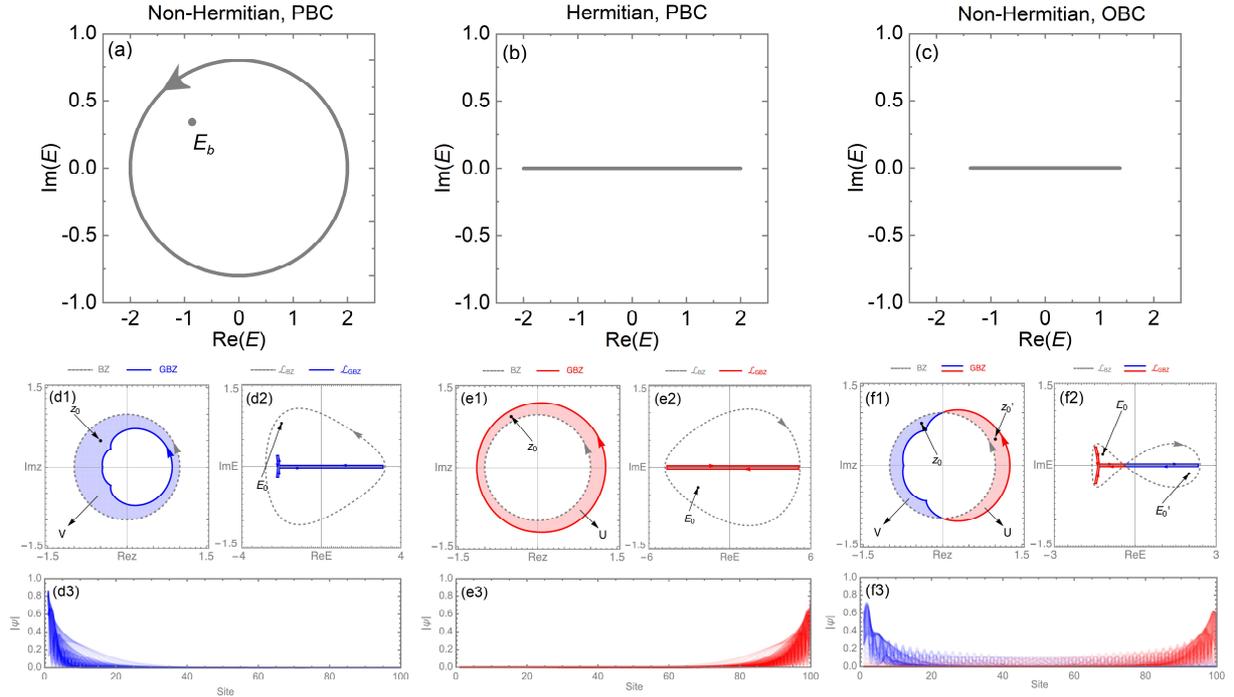

Figure 3. Energy windings under PBC for (a) the HN model with $t = 1, \gamma = 0.4$ and (b) its Hermitian counterpart. (c) The corresponding energy spectrum under OBC for the HN model in (a). (d-f) BZ and GBZ, periodic- and open-boundary spectra, together with the eigenmode distributions, for several non-Hermitian models yielding different GBZs. (d-f) are reproduced with permissions from Ref. [44].

Compared to the usual BZ with the phase factor $e^{ik}$ that encloses a unit circle as $k$ winds from $-\pi$ to $\pi$, the GBZ with $\beta$ in general does not equal to a unit cell as $r$ typically deviates from 1. The corresponding physical consequences are non-Bloch transports with directional amplifications or attenuations under OBCs, consistent with the NHSE features discussed above. Rigorous proof also indicates that while GBZ forms a closed loop similar to the conventional BZ, it exhibits completely different mathematical features. For example, the GBZ encloses the same number of zeros and poles for the Hamiltonian as holomorphic functions, such that they cancel each other [44]. This indicates no matter what base point is chosen, the eigenenergy of the Hamiltonian under GBZ winds to zero, which further suggests the trajectories of the eigenenergy under GBZ (or in other words the open-boundary spectrum of the Hamiltonian) should not form any closed loop, but instead should collapse into arcs or segments. This is consistent with the observations in the prototypical HN model under OBCs [see Figure 3(c)]. In Ref. [44], several examples that exhibit versatile NHSE phenomena yielding different GBZs were demonstrated, as adapted in Figure 3(d-f).

In addition to the relation between the nonzero energy winding under PBCs and the presence of skin modes under OBCs, the same work also suggested an interesting interpretation of the NHSE by establishing another connection stating that the nonzero winding and the NHSE have the common physical origin as the nonvanishing current through the system. If the current runs over a loop, upon truncations, the charge will start to accumulate to one end of the open system and form skin modes. If on the other hand the current runs on an arc or a segment, the winding over a closed domain (e.g., a BZ) generates zero net current and accordingly there will not be any charge accumulation upon truncations.

In the above discussions, we have reviewed three main theories that interpret the NHSE from both mathematical and physical points of view, which have provided comprehensive, consistent and complementary understanding on the NHSE. In particular, the understating on the relation between the nonzero winding of the eigenenergy of the non-Hermitian Hamiltonian under PBCs and the presence of NHSE under OBCs via the calculable GBZ technique has offered complete formulism and physical interpretation within a single theoretical scope and therefore become well-accepted, which recently has been generated to higher dimensions [102].

**2.3 NHSE enriched by symmetries**

A very important research realm for NHSE, in addition to its fundamental demonstrations, is the explorations of NHSE enriched by various symmetries. The above discussed prototypical HN model has no symmetry constraint, which already gives the interesting NHSE. Considering extra symmetries naturally brings new degrees of freedom that may interplay with the NHSE, leading to even more exciting physics and phenomena, such as the breakdown of the conventional BBC [26, 49, 57] and the defective edge states [29, 63, 145], as mentioned above and will be discussed in details in the following.

Topological phases of matter have been hot topics in condensed matter physics and their research interest is still growing, accompanying with the discoveries of novel topological phases, their robust transport behaviors and more excitingly the explorations of potential applications of the topological materials [133, 134, 146-148]. Among them, the crystalline (or lattice) symmetry-induced topological phases of matter in particular have attracted tremendous attention [149], especially in classical wave systems where artificial materials provide highly controllable and designable platforms for the realization of novel topological phases that are elusive or even absent in naturally-occurring materials [133, 134, 146-148]. The simplest example is the so-called Su-Schrieffer-Heeger (SSH) model [150], which consists of spinless particles hopping on a 1D lattice with staggered intra- and inter-cell hopping, as sketched in Figure 4(a), where the hopping amplitudes are represented by $t_1$ and $t_2$, respectively. The corresponding Hermitian Hamiltonian is written as

$$H_{\text{SSH}}(k) = \begin{pmatrix} 0 & t_1 + t_2 e^{-ik} \\ t_1 + t_2 e^{ik} & 0 \end{pmatrix}. \tag{5}$$

This model contains two sublattices of A and B and therefore obeys the chiral symmetry, which presents a $Z_2$-type topology [61]. When the inter-cell hopping is stronger, i.e., $|t_1| < |t_2|$, the system is topologically nontrivial and hosts zero-energy edge modes, one at each boundary under OBCs. When the intra-cell hopping is stronger, i.e., $|t_1| > |t_2|$, the system goes into topologically trivial region. The topological transition occurs at exactly $|t_1| = |t_2|$. Such topological properties can be identified either by tracing the energy gap-opening-closing-reopening process under PBCs [as exemplified in Figure 4(b-d)], or by searching for zero-energy edge modes under OBCs [see Figure 4(e), the red curve]. The identification of the topological phase diagram under PBCs predicts exactly the parameter regimes where the edge modes emerge under OBCs, and vice versa. This is the well-known BBC in Hermitian systems. Figure 4 (f-i) presents the mode profiles for some randomly selected eigenstates, where the extended bulk Bloch modes [see Figure 4(f)] and localized edge modes separate at the left and right chain boundaries [see Figure 4(g-i)] are clearly distinguishable.

When non-Hermiticity is induced by considering nonreciprocal hopping, however, very different behaviors are observed. Some typical signatures include the localizations for an extensive number of bulk states, the breakdown of the conventional BBC and the defectiveness of the edge modes. Based on the above analyses for NHSE, it is expected that driven by the non-Hermiticity, the originally extended Bloch bulk modes become localized skin modes. More importantly, it has been noticed that the NHSE also affects the topological properties of the SSH model, leading to a breakdown of the conventional BBC, as firstly pointed out by Yao and Wang [26] and also corroborated by some earlier observations [29, 30]. They found that the band gap closing point identified under PBCs can no longer indicate the emergence of the zero-energy edge modes under OBCs. Instead, there is a large deviation between the open boundary spectra and the periodic boundary spectra, as illustrated in Figure 4(j-n) for an exemplified non-Hermitian SSH model with nonreciprocal hopping, whose non-Hermitian Hamiltonian reads

$$H_{\text{NH-SSH}}(k) = \begin{pmatrix} 0 & t_1 + (t_2 - \gamma)e^{-ik} \\ t_1 + (t_2 + \gamma)e^{ik} & 0 \end{pmatrix}, \quad (4)$$

where the nonreciprocity is introduced on the inter-cell hopping and characterized by an offset $\gamma$. The corresponding interesting behaviors of the bulk and zero-energy edge modes driven by the NHSE in such a system are presented in Figure 4(o-r), where the bulk states become skin modes localized around the left chain boundary while the edge states experience a transition from being separate at the two chain ends to becoming identical edge skin modes localized around the left chain end. The latter corresponds to the defectiveness of the edge states and will be discussed later.

To establish a generalized BBC, dubbed non-Bloch BBC that is able to describe such non-Hermitian topological systems, Yao and Wang [26] proposed the GBZ method and based on the GBZ, they successfully derived the topological phase transition points under OBCs, which were shown to be indeed quantitively different from that under PBCs, providing concrete evidence to the breakdown of Hermitian BBC in non-Hermitian systems. Subsequently, they proposed and evaluated the non-Hermitian winding number defined on and run over the GBZ, whose results faithfully determine the non-Hermitian topological edge states and hence establish the non-Bloch BBC, reminiscent of the Hermitian BBC where the topological invariants faithfully determine the emergence of the Hermitian edge states. Note that the winding number mentioned here is defined as the winding of the *eigenstates* over the GBZ [26], different from the winding number discussed above in Equation 2, which is defined as the winding of the *eigenenergy* over the conventional BZ [44].

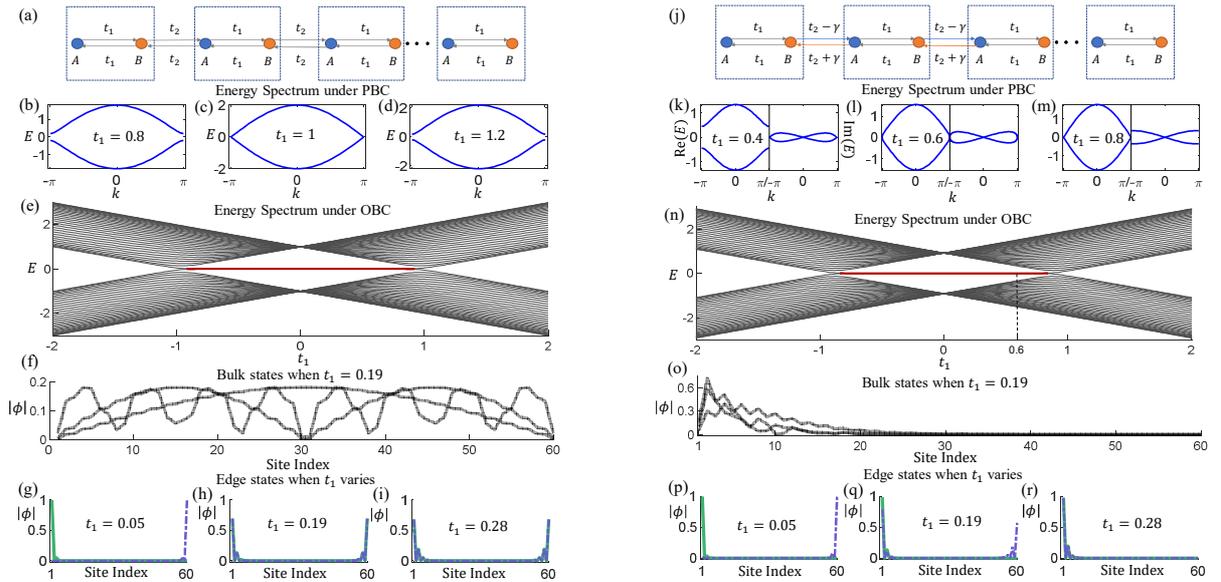

Figure 4. (a) The Hermitian SSH model, its energy spectra under (b-d) PBC and (e) OBC. (f-i) Mode profiles for the bulk and edge states. (j-r) The same as (a-i), only for the non-Hermitian SSH model. The figures are adapted with permissions from Ref. [145].

As mentioned above, the NHSE not only leads to the breakdown of the conventional BBC, but also has nontrivial effects on the topological edge states. As shown in Figure 4(o-r), driven by the non-Hermiticity, both the bulk and edge states transit into non-Hermitian skin modes. However, different from the bulk skin modes that can be observed for any nonvanishing non-Hermiticity, the edge skin modes are only found when the non-Hermitian strength is large enough. This phenomenon was identified in a recent study [145], which showed that there exists a competition between the band topology protected by certain crystalline symmetry and the NHSE, leading to an unambiguous transition from symmetry dominance to non-Hermitian dominance. In the former, the edge states are distributed separately at the two edges of the open-boundary system, which are imposed by the crystalline symmetry (e.g., the chiral symmetry). In the latter where the non-Hermiticity is strong enough, the NHSE triumphs over the protection of the crystalline symmetry and forces the edge states to transit into edge skin modes. Due to the defectiveness of the non-Hermitian Hamiltonian, the resultant edge skin modes are also defective, i.e., they collapse into a single skin mode [29,145]. This is another interesting phenomenon enabled by the interplay between NHSE and symmetries.

In addition to the above discussions, Okuma *et al.* [43] also suggested another type of symmetry-protected skin effect, i.e., the $Z_2$ NHSE, which was found to appear in the spinful systems that respect time-reversal symmetry. Similar to the quantum spin Hall effect, the time-reversal symmetry supplies a $Z_2$ invariant that enables a pair of skin modes localized at the two ends of a non-Hermitian open chain. A canonical model was proposed by combining the HN model with its time-reversed partner. It is pointed out that because the $Z_2$ NHSE is protected by the time-reversal symmetry, an infinitesimal symmetry-perturbation may break down the skin effect.

## 3. Higher-dimensional NHSE

With the flourishing of NHSE in 1D systems, it is nature to extend the related studies to two and higher dimensions, where the dimension surprisingly gives rise to versatile NHSE phenomena. In this section, we review the recent progress of NHSE in two and higher dimensions.

A direct extension of the HN model to two and higher dimensions is the simplest generalization. As schematically illustrated in Figure 5, higher dimensions give more freedom to control the nonreciprocity. For instance, if the nonreciprocity is only imposed along one direction, the NHSE leads to line and surface wave localizations respectively in two and three dimensions. If the nonreciprocity is imposed along more than one direction, 2D corner NHSE and 3D hinge and corner NHSE can be obtained, providing versatile wave manipulations.

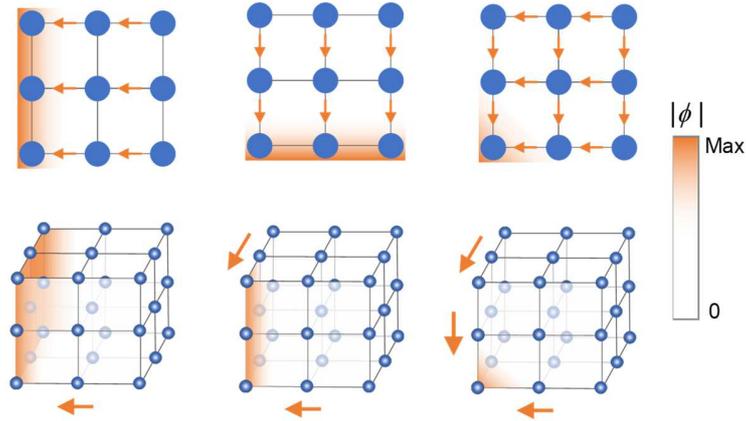

Figure 5. Higher dimensional extensions of the HN model. The orange shading schematically represents the distributions of the skin modes. Note that only the primary hopping is indicated for the sake of clarity.

In HN-like models, there is no symmetry constraint and hence all eigenstates are skin modes, i.e., the skin modes scale with $L^d$ where $L$ denotes the length of the finite lattice with OBCs and $d$ represents the dimension. When combined with symmetries, however, the symmetries impose restrictions on the hopping particles and accordingly the skin modes do not necessarily scale with $L^d$, but instead exhibit versatile varieties. In Ref. [92], Lee *et al.* combined the nonreciprocity with chiral symmetry in a 2D square lattice consisting of four hopping sites in each unit cell [see Figure 6(a)]. The nonreciprocity is considered in both *x*- and *y*-directions with different threading, such that by manipulating the threading, the net nonreciprocity of the whole lattice can be skillfully controlled. It was shown that for a general threading, corner NHSE similar to that in a 2D HN model is observed, as shown in Figure 6(b). If the nonreciprocity is destructively canceled along the *x*-direction, NHSE only appears in the *y*-direction [see Figure 6(c)]. When the nonreciprocity is destructively canceled along both directions and hence the net nonreciprocity vanishes, counterintuitively, corner skin modes were again observed under full OBCs, which were found to be the consequences of the 1D topological edge states experiencing local nonreciprocity [see Figure 6(d) for the topological edge states and Figure 6(e) for the corner skin modes]. Considering their origin, these skin modes are dubbed hybrid skin-topological modes and they scale with $L$ [92]. The similar results have been also generated to three dimensions. Later, a more comprehensive understating on the skin-topological properties was reported [98], providing analytical deductions and interpretations for different localization behaviors.

Following this direction, others non-Hermitian lattices yielding NHSE were also proposed. For example, taking the Benalcazar-Bernevig-Hughes model [151] as the extended Hermitian Hamiltonian, the resultant non-Hermitian model enables an interesting NHSE that hosts corner skin modes scaling with $L$ while accompanying $\mathcal{O}(L^2)$ extended bulk modes [95]. In a later work [99], based on the model in Ref. [95] and by adding on-site disorders, an energy-asymmetrical NHSE can be obtained where in some regions in the complex energy plane, the corner NHSE manifests while in other regions, the NHSE disappears due to the disorder localizations.

Note that the terminology higher-order NHSE has been widely, which, however, refers to different NHSE phenomena in different literature and does not have a universal definition. Hence, to avoid possible confusion, in this review, we do not refer to higher-order NHSE for any particular NHSE phenomenon, but simply use it as a terminology indicating the general NHSE phenomena in higher dimensions.

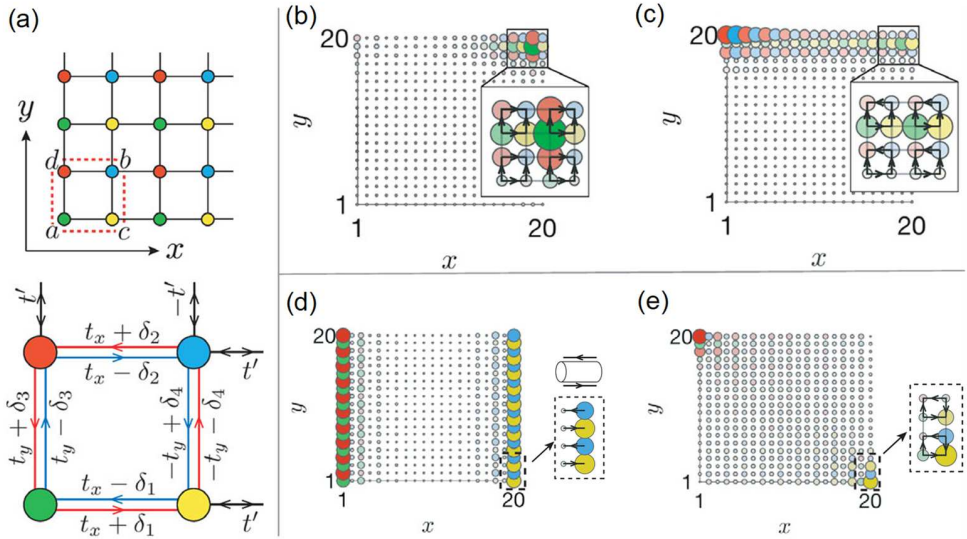

Figure 6. (a) The 2D tight-binding model, where NHSE is manifested as (b) corner skin modes for nonreciprocity in both directions and as (c) edge skin modes when nonreciprocity is canceled in the *x*-direction. (d-e) Hybrid skin-topological modes with nonreciprocity canceled in both directions. The figures are reproduced with permissions from Ref. [92].

Another phenomenon enabled by NHSE, known as the non-Bloch PT symmetry, has been briefly discussed in Section 2.2, associated with the nonreciprocity-induced EPs. Generating to higher dimensions, novel features arise simply due to the increasing of spatial dimensions [138]. Specially, it was shown that the non-Bloch PT phase breaking in two and higher dimensions universally approaches to zero as the system size increases, whereas the Bloch PT phase breaking and the 1D non-Bloch PT phase breaking generally have nonzero thresholds. Although a physical explanation is still lacking, this interesting phenomenon suggests rich and unexpected interplay among PT symmetry, NHSE, and spatial dimensions.

When versatile NHSE phenomena have been reported, the contemporary theories for higher-dimensional NHSE have been also explored. In Ref. [102], Zhang *et al.* proposed a theorem that relates the emergence of NHSE in two and higher dimensions to the periodic-boundary spectrum of the non-Hermitian Hamiltonian by requiring the latter to cover a finite area on the complex energy plane, similar to the NHSE theorem in 1D systems where the periodic-boundary spectrum is required to form loops (see detailed discussions in Section 2.2). While providing a phenomenological indication to the higher-order NHSE, this theorem does not give a rigorous formulism as that in 1D case using the GBZ technique. In fact, it has been shown that the GBZ theory cannot be generated to higher dimensions [44]. An alternative was proposed via numerical calculating the topological invariant based on the eigenstates under OBCs [53]. When consistent with GBZ theory in 1D cases, this method can be generalized to two dimensions. However, it has been only applied to topological systems and its applicability to higher-order NHSE phenomena still requires verification. In Ref. [95], the Wess-Zumino term was introduced to specifically explain the emergence of the corner skin modes in the non-Hermitian lattice under study. Overall, it is fair to say that the theories for higher-order NHSE are somewhat case by case, leaving the general formulism still an open question.

## 4. NHSE with extra physical degrees of freedom

In addition to generating the NHSE phenomena to higher dimensions, another research trend is to explore the interplay between NHSE and extra physical degrees of freedom such as long-range coupling, pseudospins, magnetism, non-linearity, crystal defects, etc., which will be reviewed in this section.

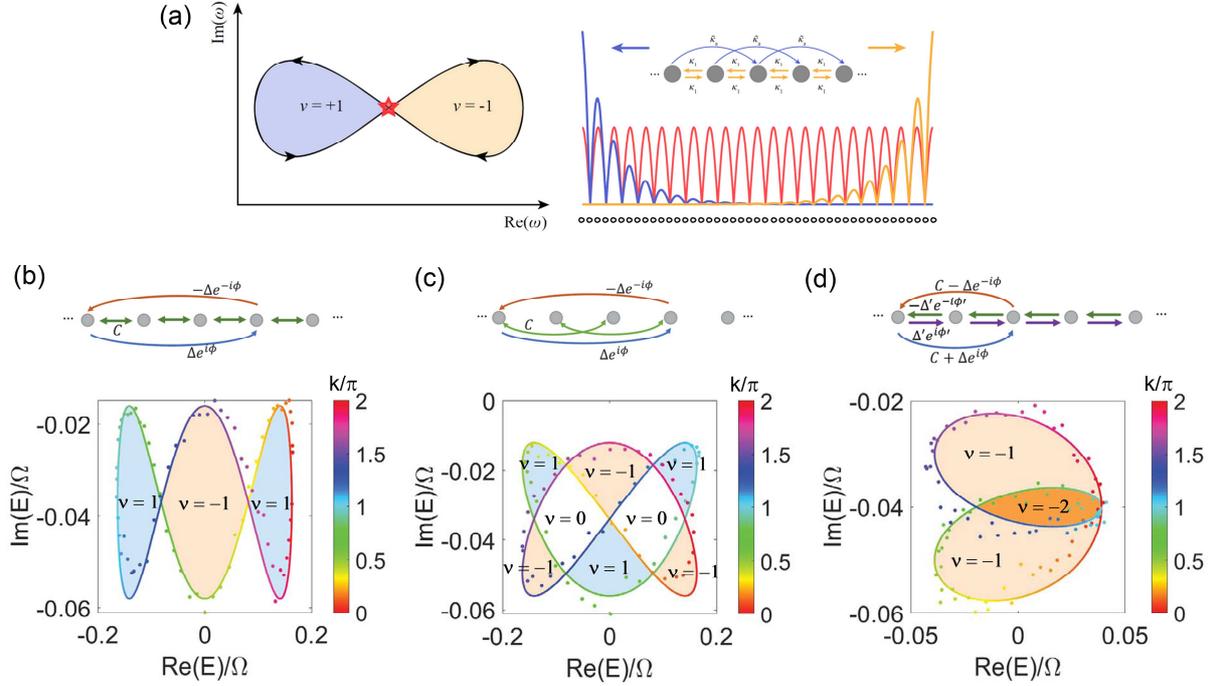

Figure 7. (a) Twisted windings in the HN model with long-range coupling and its bipolar skin effect. (b-d) Various windings by manipulating the long-range couplings. The figures are reproduced with permissions from Refs. [108, 117].

First, we show how the long-range coupling can be a powerful tool to enrich the nonreciprocity and accordingly lead to versatile NHSE phenomena. In Ref. [53], Song *et al.* proposed that by introducing long-range coupling, a bipolar NHSE can be obtained where the skin modes appear either at the left or the right chain end depending on its eigenenergy. In a later work [117], the HN model with long-range coupling was experimentally realized to demonstrate the bipolar NHSE in an acoustic system. It was shown that this effect is essentially due to the generating of two regions of opposite eigenenergy windings [117], as illustrated in Figure 7(a). A more comprehensive study even suggested it is possible to generate arbitrary windings by manipulating the long-range couplings (see Figure 7(b-d) for adapted examples from Ref. [108]).

Besides the long-range coupling, there are many other ingredients that make the NHSE flavorful. For instance, a recent study exploited the pseudospin degrees of freedom emulated by whisper-gallery modes in ring waveguides and realized the spin-dependent NHSE in a passive system [114]. Another work identified that the magnetic fields can suppress the NHSE, leading to the Onsager-Lifshitz quantization rule persisting in the long-wavelength limit regardless of the NHSE [152]. Meanwhile, a real-to-complex transition of the energy spectra was found to appear in such a system upon the spontaneous breaking of an underlying mirror-time reversal symmetry, a reminiscent of the PT symmetry and its phase transition. More recently, the non-linearity has also been introduced to the studies of NHSE, which was shown to be able to realize NHSE with symmetric couplings [109] or to dynamically modify the structures of the skin modes [153]. Another interesting exploration of NHSE with flavor is with the assistance of crystal defects such as dislocations [116,118] and disclinations [111]. It was shown that driven by the carefully-designed nonreciprocity, an extensive number of eigenstates become localized at the vicinity of the dislocations or disclinations, exhibiting novel features for NHSE that does not require and can be disentangled from the open boundaries. Note that due to the geometric nature of the crystal defects, the defect-induced NHSE phenomena are unique in two and higher dimensions.

## 5. Experimental progress

Accompanying the theoretical developments, the recent years have also witnessed the experimental progress for NHSE. In this section, we review the experimental platforms, various ways to introduce nonreciprocity, the experimental characterizations of the NHSE, as well as the demonstrated special and unique NHSE features and phenomena.

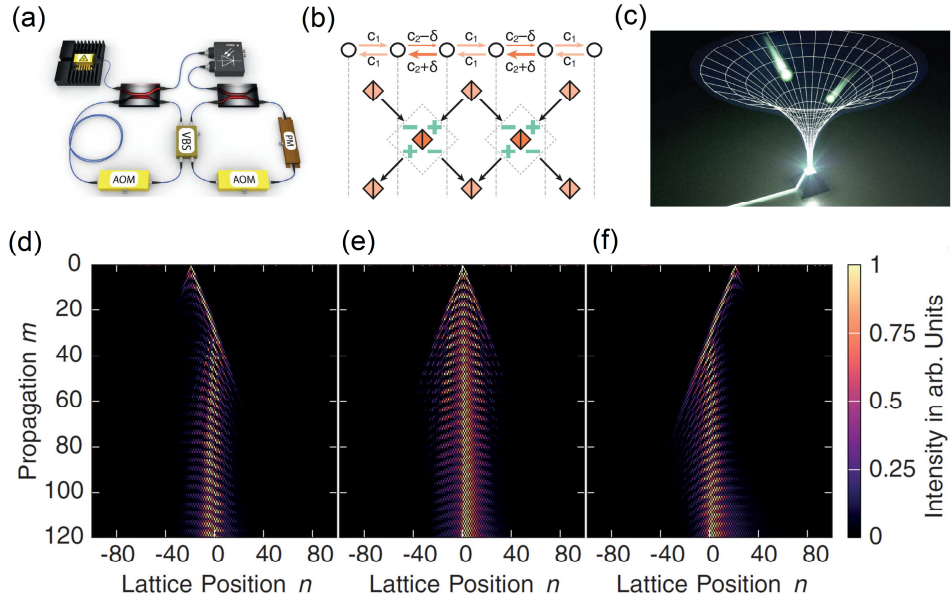

Figure 8. (a) The experimental set-up of optical fiber loops. (b) Time-modulation to realize nonreciprocal couplings. (c) A schematic light funnel proposed using the NHSE, whose experimental measurements are shown in (d-f) for different input of light. The figures are reproduced with permissions from Ref. [121].

As one of the earliest experimental demonstrations of the NHSE, Ref. [121] used a 1D photonic lattice consisting of optical fiber loops, whose set-up is shown in Figure 8(a). The amplitudes of the left- and right-moving photons are controlled by a time-modulation on their hopping paths via a beamsplitter that essentially mediates the hopping strength between the lattice sites [see Figure 8(b)]. When the mediation is neutral, the photonic lattice behaves as an SSH lattice supporting topological edge states. When the mediation becomes biased, the system becomes nonreciprocal and the NHSE is manifested, with all eigenstates exponentially localized at the interface terminated by another photonic lattice with opposite mediation. Based on such an NHSE phenomenon, the authors proposed a light funnel that can efficiently guide any incident light toward the designed interface, irrespective of the shape and input positions of the incidence, as schematically illustrated in Figure 8(c) and experimentally verified in Figure 8(d-f).

The similar time-modulating technique for photons was also used in the field of non-unitary quantum-walks to realize the nonreciprocal hopping [see Figure 9(a)] and hence the 1D NHSE [122]. In the same work, the interesting non-Hermitian BBC was further demonstrated by directly observing the non-Hermitian topological edge states and their agreement with the non-Bloch topological invariants defined on the GBZ.

Another implementation of the nonreciprocity-induced non-Hermiticity in photonics was facilitated in the scope of synthetic dimensions, where the multiple frequency modes in a ring resonator are treated as a 1D synthetic dimension [108]. The couplings among the modes are controlled by the amplitude and phase modulations, with the nonreciprocity induced by detuned modulations. Using this set-up, the authors have demonstrated arbitrary non-Hermitian windings in a HN-like model with long-range couplings, as discussed in Section 4. The set-up for this design is illustrated in Figure 9(b) for the ring resonator (left panel) and the synthetic frequency dimension (right panel).

On another front, the electric circuits have been proven to be very powerful in mimicking the tight-binding models, as the industrially well-refined electric components are capable of accommodate exact and almost arbitrary control of hopping [128]. By using electric circuits, the NHSE has been demonstrated not only in 1D systems [123, 124, 126], but also in two and higher dimensions [128, 130]. The corresponding set-ups for these electric circuits are displayed in Figure 9(c-f). Due to this high controllability, electric circuits have become the most popular experimental platform for the NHSE studies. A recent preprint even suggested electric circuits can mimic a strongly correlated non-Hermitian many-body system which supports a non-Hermitian many-body skin effect [127].

While lacking the flexibility of versatile modulations and coupling control, the acoustic community has also skillfully proposed feasible techniques to experimentally observe and explore novel NHSE phenomena. For example, in Ref. [117], the acoustic nonreciprocity was realized with a directional amplifier enabled by a DC power supply, whose set-up is shown in Figure 9(g). Using this set-up, the twisted non-Hermitian windings and bipolar skin effect were reported, as discussed in Section 4.

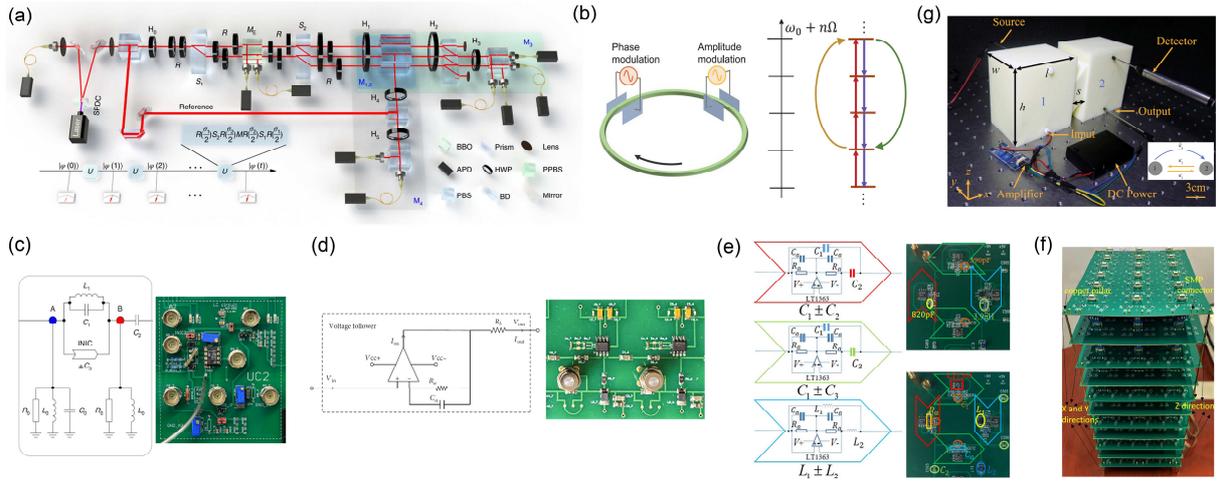

Figure 9. Experimental set-ups for NHSE demonstrations in various physical platforms, including (a) quantum-walks, (b) photonic synthetic dimensions, (c-f) 1D, 2D and 3D electric circuits, as well as (g) acoustic cavities equipped with electrical amplifiers. These figures are reproduced with permissions from Ref. [108, 117, 122, 123, 126, 128].

Other experimental studies for NHSE also include clever designs of nonreciprocity in mechanics [125, 154-156] and recently in ultracold atoms [129]. In the former examples, active mechanical components, such as local control loops with biased strain-dependent forces [125, 154, 155] and odd micropolar materials constructed with piezoelectric elements whose cycle feed induces asymmetric bending and shearing [156], were used to break reciprocity. Accordingly, interesting active and non-Hermitian features were reported, such as odd elasticity [157] and mechanical robots [155]. In the ultracold-atom experiment, the nonreciprocity was obtained by a momentum-lattice engineering method which generates a synthetic magnetic flux via the Raman or Bragg transitions. In this path, the condensed atoms experience a laser-induced dissipation process, while in the opposite path, the Raman-Bragg coupling is turned off and no dissipation occurs, hence generating nonreciprocity [129]. NHSE phenomenon was accordingly observed.

In addition to the above discussed active and additive control of couplings, a recent experimental work suggested that in a complete reciprocal system with time-reversal symmetry, it is also possible to realize NHSE [114]. As shown in Figure 10(a), the system was implemented in a 2D ring-resonator design, which supports decoupled clockwise and anticlockwise modes. A carefully designed loss configuration was introduced to the link rings, such that for each of the whisper-gallery modes, the propagation is nonreciprocal [see Figure 10(b) for the illustration in the $x$-direction and similar analyses can be performed in the $y$-direction]. But for the entire lattice, it maintains the reciprocity, i.e., for each clockwise mode propagation, there is always a time-reversed anticlockwise partner. With this design, the authors demonstrated a $Z_2$ NHSE in two dimensions, accompanying clockwise and anticlockwise corner skin modes localized at opposite geometric corners [see Figure 10(c), compared with the Hermitian counterpart in Figure 10(d)]. This was also dubbed pseudospin-dependent NHSE, as discussed in Section 4. It is pointed out that although this work was implemented in acoustics but can be readily extended to photonics where the optical resonator waveguides have been proven to be mature material platforms.

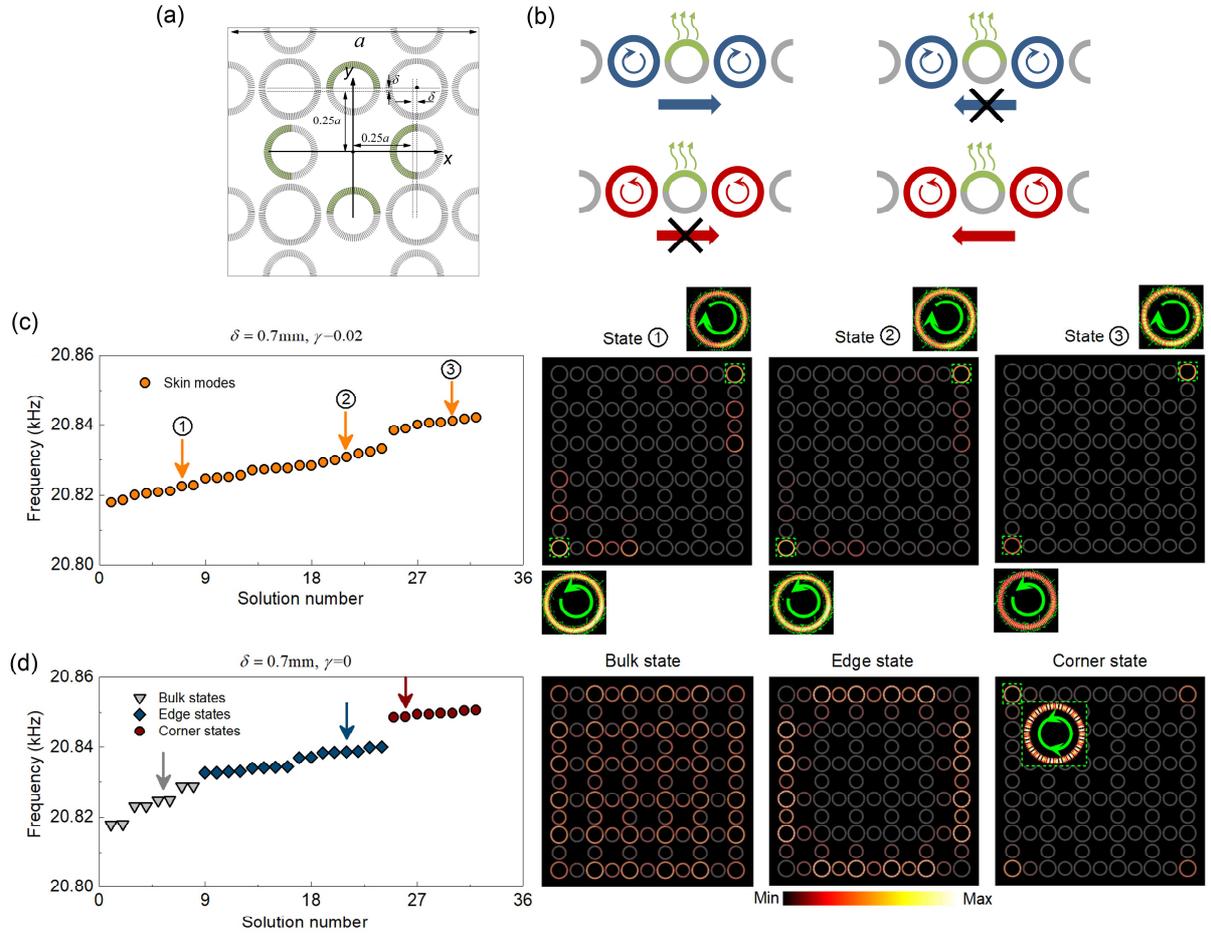

Figure 10. A unit cell of the 2D reciprocal lattice. The green areas represent the lossy domains. (b) Nonreciprocal propagation for each clockwise or anticlockwise mode. (c) Higher-order NHSE manifested as pseudospin-dependent corner skin modes. (d) The Hermitian case as the comparison. (a, c-d) are reproduced from Ref. [114].

## 6. Conclusions and outlooks

To summarize, we have reviewed the recent developments of NHSE, by illustrating in details the minimal model, the fundamental theories and physics, its interplay with symmetries, extensions to higher dimensions, the enrichment with various degrees of freedom, as well as the experimental progress. We believe such a comprehensive, informative and timely review on this fast-growing filed will not only provide hands-on practice for junior researchers, but also give rich perspectives for NHSE-related experienced researchers.

The directions to move forward could potentially rest on looking for novel NHSE phenomena in two and higher dimensions, especially enriched by various crystalline symmetries including both point-group and space-group symmetries, while the latter might require extra effort to realize nonreciprocity. Therein, topology as a flavorful ingredient for NHSE may bring even more exotic non-Hermitian physics and phenomena. Explorations on nonreciprocity-induced PT symmetry, mirror-time symmetry [152] and related exceptional features also fall into this category, which may facilitate the understanding of NHSE in higher dimensions and inspire new research that is unique to higher dimensions. Other topics such as incorporating various physical degrees of freedom including active components (which already have been widely explored), gauging control, synthetic features, etc., exploring other experimental platforms for both nonreciprocal and reciprocal realizations of NHSE, are also possible directions.

Conclusively, while many of the reviewed results, for the most part, have contributed to the understanding of basic physics and can only be categorized in terms of academic research or proofs of concepts, based on the unconventional ways various particles are manipulated, there have been already many convincing suggestions to potential applications such as unidirectional amplifiers [158], high-efficient energy harvesting via an edge burst [159] and autonomous mechanical robots [155]. With these visions, one can only look forward to unprecedented routes and possibilities for technologies that take advantage of the novel non-Hermitian and nonreciprocal physics.

31